\documentclass[aps,twocolumn,superscriptaddress]{revtex4-2}
\usepackage[utf8]{inputenc}
\usepackage{graphicx}
\usepackage{amsmath,color}
\usepackage{natbib}
\usepackage{amsfonts}
\usepackage{bbold}
\usepackage{nicefrac}
\usepackage{bm}
\usepackage{ragged2e}
\usepackage{mathtools}
\usepackage{hyperref}
\usepackage{braket}
\hypersetup{
    colorlinks=true,
    linkcolor=blue,
    filecolor=mangeta,
    citecolor=blue,
    urlcolor=cyan
}
\begin{document}
\author{Luan M. Ver\'{i}ssimo}
\affiliation{Instituto de F\'{i}sica, Universidade Federal de Alagoas 57072-900 Macei\'{o} - AL, Brazil}
\affiliation{Donostia International Physics Center, Paseo Manuel de Lardizabal 4, E-20018 San Sebasti\'{a}n, Spain}
\author{Marcelo L. Lyra}
\affiliation{Instituto de F\'{i}sica, Universidade Federal de Alagoas 57072-900 Macei\'{o} - AL, Brazil}
\author{Rom\'{a}n Or\'{u}s}
\affiliation{Donostia International Physics Center, Paseo Manuel de Lardizabal 4, E-20018 San Sebasti\'{a}n, Spain}
\affiliation{Multiverse Computing, Paseo de Miram\'{o}n 170, E-20014 San Sebasti\'{a}n, Spain}
\affiliation{Ikerbasque Foundation for Science, Maria Diaz de Haro 3, E-48013 Bilbao, Spain}
\title{Dissipative Symmetry-Protected Topological Order}

\begin{abstract}
In this work, we investigate the interplay between dissipation and symmetry-protected topological order. We considered the one-dimensional spin-1 Affleck-Kennedy-Lieb-Tasaki model interacting with an environment where the dissipative dynamics are described by the Lindladian master equation. The Markovian dynamics is solved by the implementation of a tensor network algorithm for mixed states in the thermodynamic limit. We observe that, for time-reversal symmetric dissipation, the resulting steady state has topological signatures even if being a mixed state. This is seen in finite string-order parameters as well as in the degeneracy pattern of singular values in the tensor network decomposition of the reduced density matrix. We also show that such features do not appear for non-symmetric dissipation. Our work opens the way toward a generalized and more practical definition of symmetry-protected topological order for mixed states induced by dissipation.  
\end{abstract}

\maketitle

{\it Introduction.-} Topological phases of matter \cite{Top1,Top2} are distinct quantum phases with {short-range entanglement and signature patterns in entanglement spectrum} \cite{Hal}. Some of such phases are also protected by existing symmetries in the system, being thus called symmetry-protected topological (SPT) phases. An archetypal example is the 1d Affleck-Kennedy-Lieb-Tasaki(AKLT) model  \cite{AKLT}, where $Z_2 \otimes Z_2$ and time-reversal symmetries protect the structure of entanglement giving rise to non-zero string-order parameters  \cite{VBS2}, spin-1/2 edge modes  \cite{Hal}, and topological degeneracy of the entanglement spectrum  \cite{ESHaldane}, which survive under the action of symmetry-preserving perturbations  \cite{Pollman1,Pollman2}. The presence of edge modes prevents such topological phases from being continuously deformed into a topologically trivial phase without going through a quantum phase transition, where the edge modes penetrate the bulk and the energy gap of the system closes. 

An important problem that gain attention in recent years is how to combine effectively topological order and dissipation \cite{TD1,TD2,TD3,TD4}. This is due to the emerging huge potential of dissipative effects in quantum computing \cite{iTQC,TD4} and material science \cite{Top3,Top4}. Recently, a definition of topological phases of matter for mixed-state was proposed \cite{oSPTP2} in the light of fast dissipative evolution with local and time-independent symmetric Lindbladians. With that definition, the authors demonstrate that SPT order is destroyed under dissipation. Another recent work \cite{oSPTP} introduces the conditions to preserve the SPT order under specific types of dissipations. The so-called strong symmetry condition for dissipative dynamics is satisfied when the local jump operators commute with the operators composing the projective representation of the symmetry group of interest and also with the Hamiltonian of the system. This condition guarantees the preservation of the string order for SPT under dissipation satisfying the strong symmetry condition \cite{oSPTP}.

In this work, we address the interplay between dissipation and SPT order from a different angle. In particular, we consider the action of symmetric dissipation on a 1d SPT phase and evaluate it in terms of mathematical quantities such as the degeneracy of the Schmidt coefficients of the tensor network representation of the mixed state. In the case of pure states, such Schmidt coefficients relate directly to the entanglement spectrum of the state. In mixed states, however, such coefficients do not have such a straightforward physical interpretation, but as we shall see, offer also a valid mathematical tool to assess long-range patterns of correlations (classical and quantum) in the mixed state. 

In particular, we consider the 1d AKLT model interacting with an environment, where the Markovian dynamics is given by the Lindblad master equation. The dissipation operators are chosen in order to preserve (symmetric) or break (non-symmetric) time-reversal  symmetry. We show, via numerical tensor network simulations, how these two cases lead to steady states with very different properties in terms of long-range correlations and other typical signatures of SPT order. 

\vspace{3pt}

{\it Model and method.-} In order to investigate the effects of symmetric and  non-symmetric dissipation on SPT order we considered an interacting quantum spin model given by the Spin-1 AKLT Hamiltonian,  

\begin{equation}
    H =  \sum_{i} \vec{S}_{i} \cdot \vec{S}_{i+1} + \frac{1}{3} \left(\vec{S}_{i} \cdot \vec{S}_{i+1} \right)^2, 
\end{equation}
where $\vec{S}$ represents the three-component spin-1 operator $\vec{S} = \left(S_x,S_y,S_z \right)$, for an infinite chain with open boundary conditions. The above Hamiltonian has $SU(2)$ symmetry, and the ground state is given by the so-called AKLT state. Such order has a structure of entanglement protected by $Z_2 \otimes Z_2$ symmetry  \cite{AKLT,Kennedy}, and defines a topological Haldane phase. Our strategy here is to couple the AKLT model to an environment, time-evolve the new physical system until reaching the steady state and assess the signatures of topological order. The dissipative evolution follows a master equation of the type  
\begin{equation} \label{ME}
    \dot \rho = \mathcal{L} \left[ \rho \right] = -i \left[ H, \rho \right] + \gamma \sum_{\mu} \left( L_{\mu} \rho L^{\dag}_{\mu} - \frac{1}{2} \left\{L^{\dag}_{\mu}L_{\mu}, \rho \right\} \right),  
\end{equation}
where $\rho$ is the density matrix of the system, $H$ the Hamiltonian of the model, $\gamma$ the coupling between the quantum system and the environment, and the pair $ \{ L_{\mu}, L^{\dag}_\mu \}$ are Lindblad jump operators. We will focus on different local choices of operators $L_\mu$, either preserving symmetries or not, and their effect on SPT order in the resulting steady state.   

\begin{figure}
	\centering
	\includegraphics[width=0.8\linewidth]{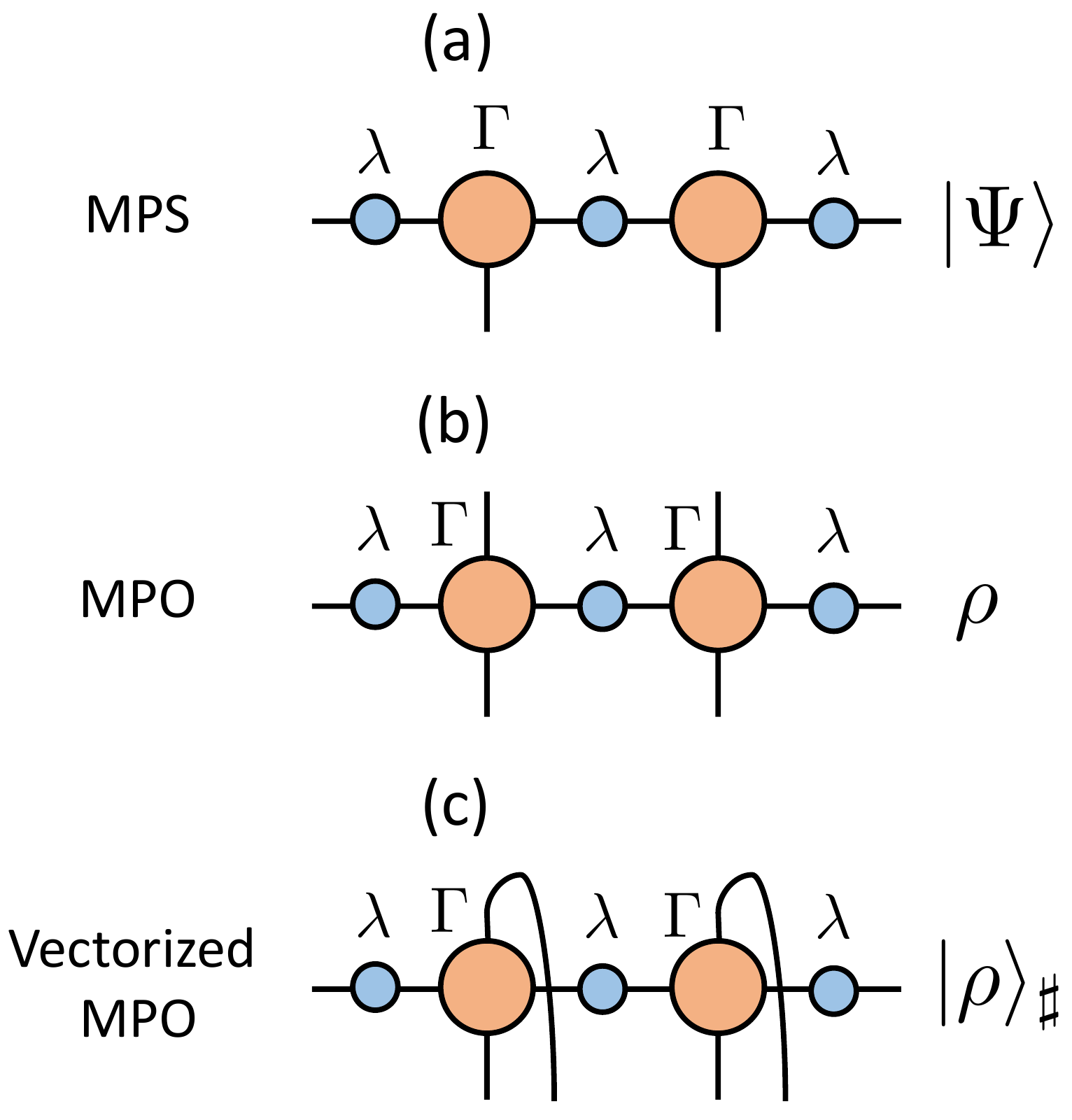}
	\caption{[Color online] Tensor network diagrams for a 1d system with a 1-site unit cell: (a) MPS for a pure state $\ket{\Psi}$, (b) MPO for a mixed state $\rho$, and (c) vectorized MPO (following Choi's isomorphism) for vectorized mixed state $\ket{\rho}_\sharp$. In the diagrams, shapes represent tensors, and lines represent indices in the tensors. Lines from one shape to another represent contracted common indices between tensors. The diagonal $\lambda$ tensors correspond to the (real and positive) Schmidt coefficients of sequential bipartitions of the tensor network and are the ones associated with the spectrum $\xi_\alpha = -2 \log ( \lambda_{\alpha})$, which amounts to the entanglement spectrum in the pure-state case represented in (a).} 
	\label{fig0}
\end{figure}

In order to study the dissipation effects on the steady states, we compute quantities such as the string order parameter $\mathcal{O}^{\alpha}_{S}$, 

\begin{equation}
  \mathcal{O}^{\alpha}_{S} = \lim_{|r| \rightarrow \infty} \left\langle S^{\alpha}_{i} e^{i \pi \sum_{l=1}^{j-1}S^{\alpha}_{l}} S^{\alpha}_{j} \right\rangle,  
\end{equation}
with $S^\alpha$ the $\alpha$-component of spin variable($\alpha=x,y,z$) and $r = i - j$. In addition, the entanglement spectrum \cite{ESHaldane,Pollman2} and its degeneracy are also important indicators of 1d SPT phases. In the non-dissipative case, this can be assessed by checking the degeneracy of the Schmidt values in a Matrix Product State (MPS) decomposition of the pure state, or equivalently, the degeneracy of the entanglement energies $\xi_\alpha \equiv -2\log(\lambda_{\alpha})$,  with $\lambda_\alpha$ the Schmidt coefficients of sequential Schmidt decompositions of the MPS. In the case of pure states, such Schmidt coefficients have a clear interpretation in terms of the eigenvalues of the reduced density matrix of half an infinite system and are therefore directly related to the quantum entanglement present in the pure state. However, in the dissipative case, the situation is slightly different. As we shall see, we shall represent the mixed state of the system using a Matrix Product Operator (MPO). Such MPO also admits a representation in terms of sequential singular value decompositions (equivalent to the Schmidt decompositions in the case of pure states), which in fact is nothing but the canonical form of the 1d tensor network  \cite{OrusiTEBD}, see Fig.\ref{fig0}. The corresponding Schmidt coefficients in the dissipative case do \emph{not} represent formally the quantum entanglement in the system since classical and quantum correlations are intertwined in this case. But we can say, however, that they represent the relevant correlation parameters of the MPO tensor network, even if their physical interpretation is not so straightforward as in the pure state case. We will see in what follows that the degeneracy patterns of these dissipative Schmidt coefficients are also relevant to assess possible signatures of long-range correlations in mixed states.     

We have performed numerical calculations using tensor network methods, which we now briefly describe. In recent years, we have witnessed a fast development of tensor networks \cite{Orusrev,DMRG1,Verstraete21} for simulations of quantum many-body systems. Methods such as \emph{time-evolving block decimation} (TEBD) \cite{Vidal1,Vidal2,OrusiTEBD} and \emph{density matrix renormalization group} (DMRG) \cite{DMRG1,DMRG2} allow exploration of ground state properties of 1d systems using MPS representations of quantum states, even in the thermodynamic limit. Extensions of tensor network methods for open quantum systems have recently been presented \cite{TNOQS,RomanRev2}. People have introduced relevant concepts, such as matrix product operators (MPO), extending the notion of MPS from pure states to mixed states in the case of 1d systems. Along these lines, the primordial work of Zwolak and Vidal \cite{Zwolak} extended the TEBD algorithm to effectively simulate the dynamics of mixed states, by solving the Lindblad master equation in vectorized form. This method was also recently implemented for two-dimensional systems \cite{Orus2d} with success. Additionally, other recent strategies \cite{Banuls} propose variational methods as well hybrid algorithms combining real and imaginary time for Lindblad evolutions \cite{hybrid} in order to find dissipative steady states. 

\begin{figure} 
    \includegraphics[width=85mm, clip]{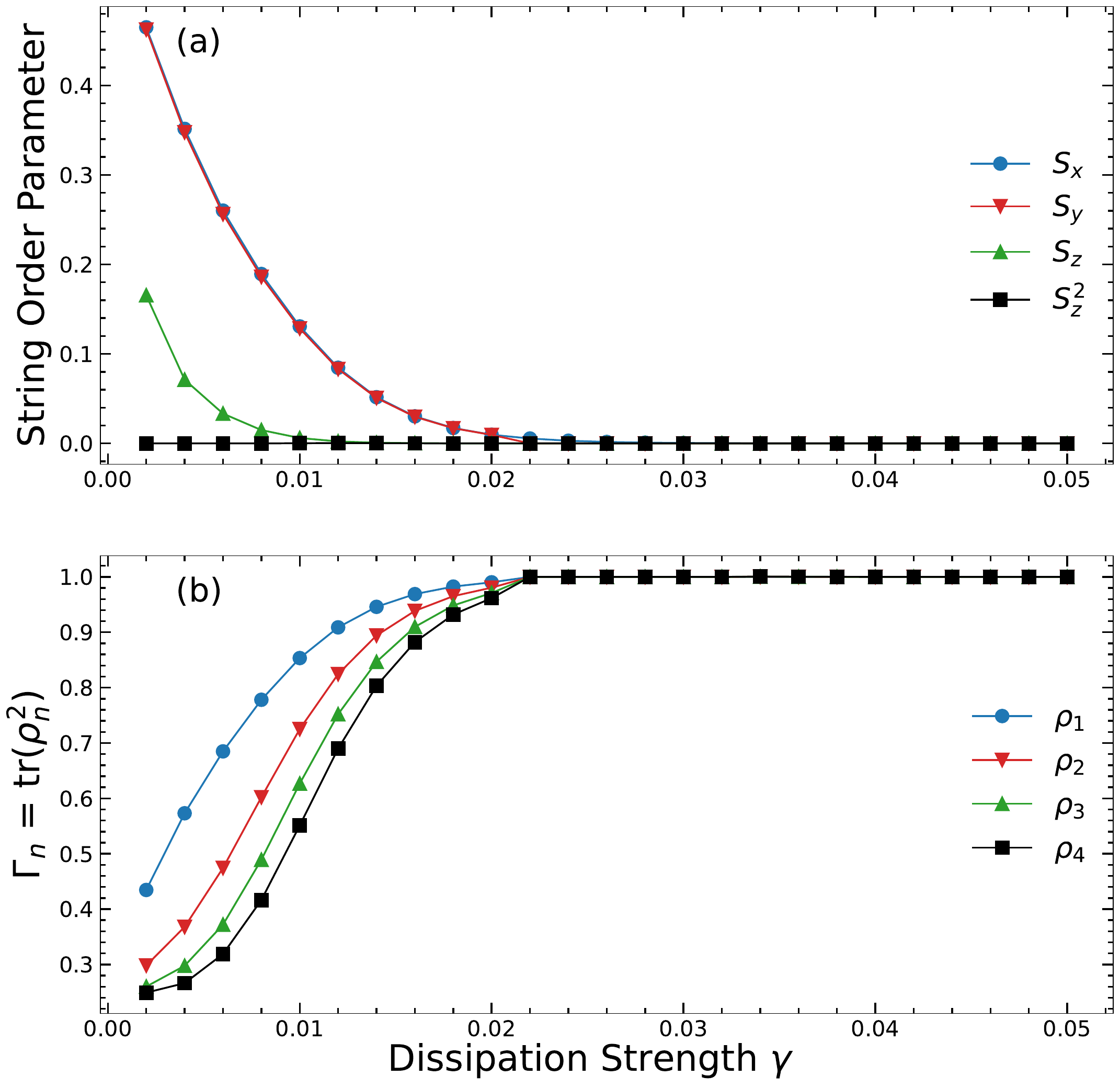}
    \caption{[Color online] (a) String Order Parameters  $\mathcal{O^{\alpha}_S}$ and (b) Purities $\Gamma_n$ for non-symmetric dissipation $L_\mu = S_z$, for different values of strength dissipation $\gamma$.} 
    \label{figA1}
\end{figure}
\begin{figure}
    \includegraphics[width=85mm, clip]{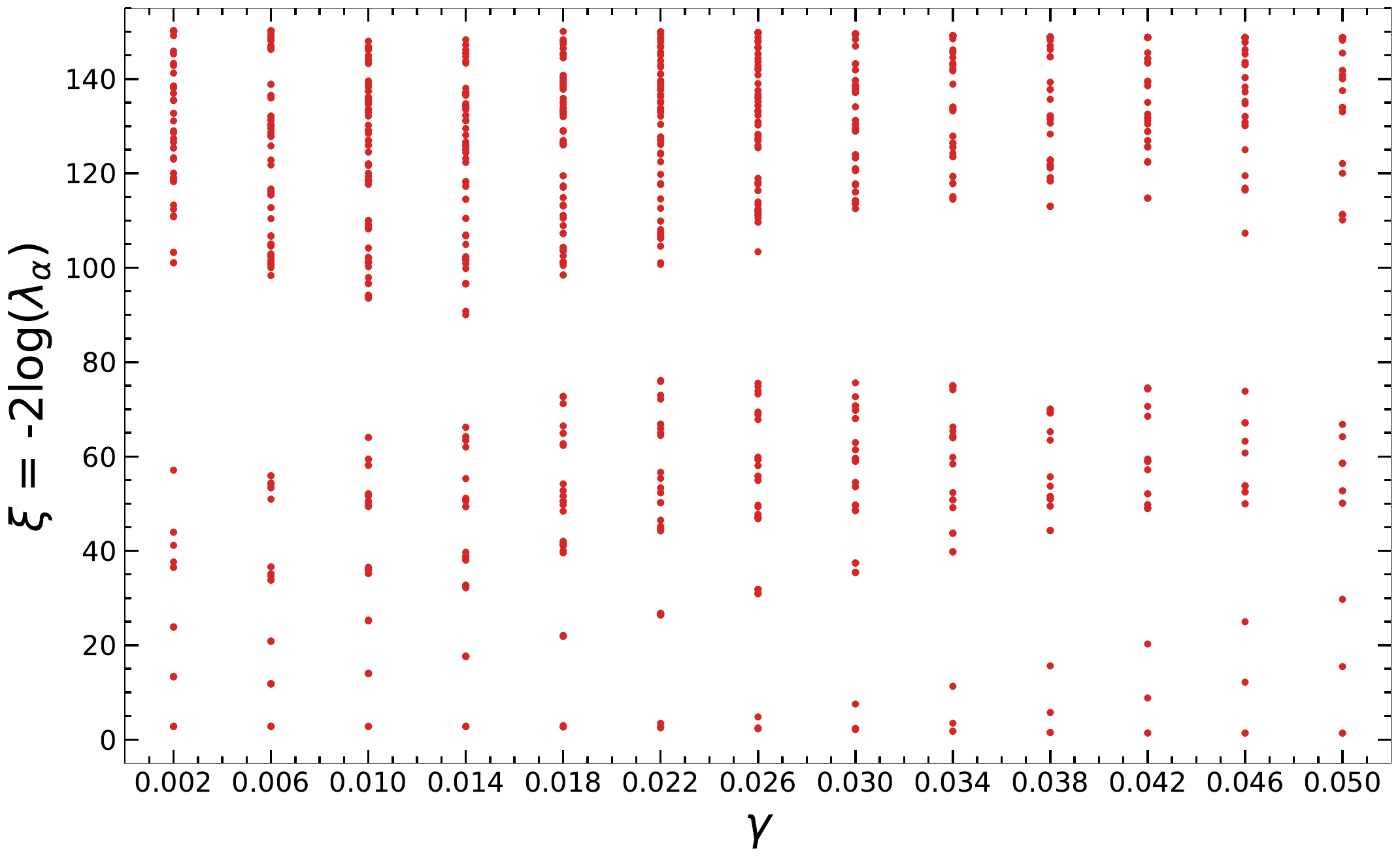}
    \caption{[Color online] Lowest part of the spectrum $\xi_\alpha$ for non-symmetric dissipation $L_\mu = S_z$, as a function of strength dissipation $\gamma$.}
    \label{figA2}
\end{figure}

\begin{figure}
    \includegraphics[width=85mm, clip]{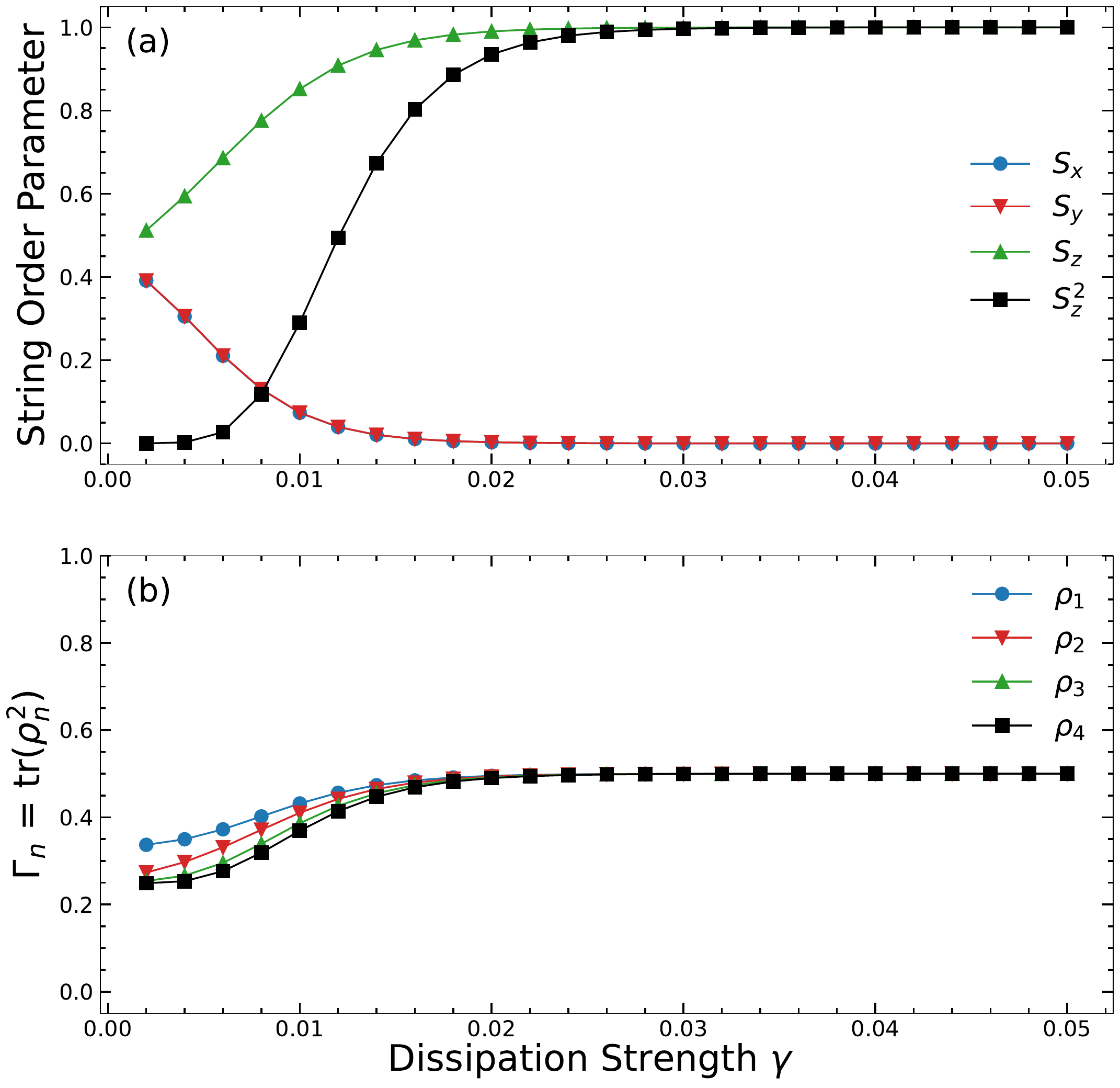}
    \caption{[Color online] (a) String Order Parameters  $\mathcal{O^{\alpha}_S}$ and (b) Purities $\Gamma_n$ for symmetric dissipation $L_\mu = S_z^2$, for different values of strength dissipation $\gamma$.}
    \label{figB1}
\end{figure}
\begin{figure}
    \includegraphics[width=85mm, clip]{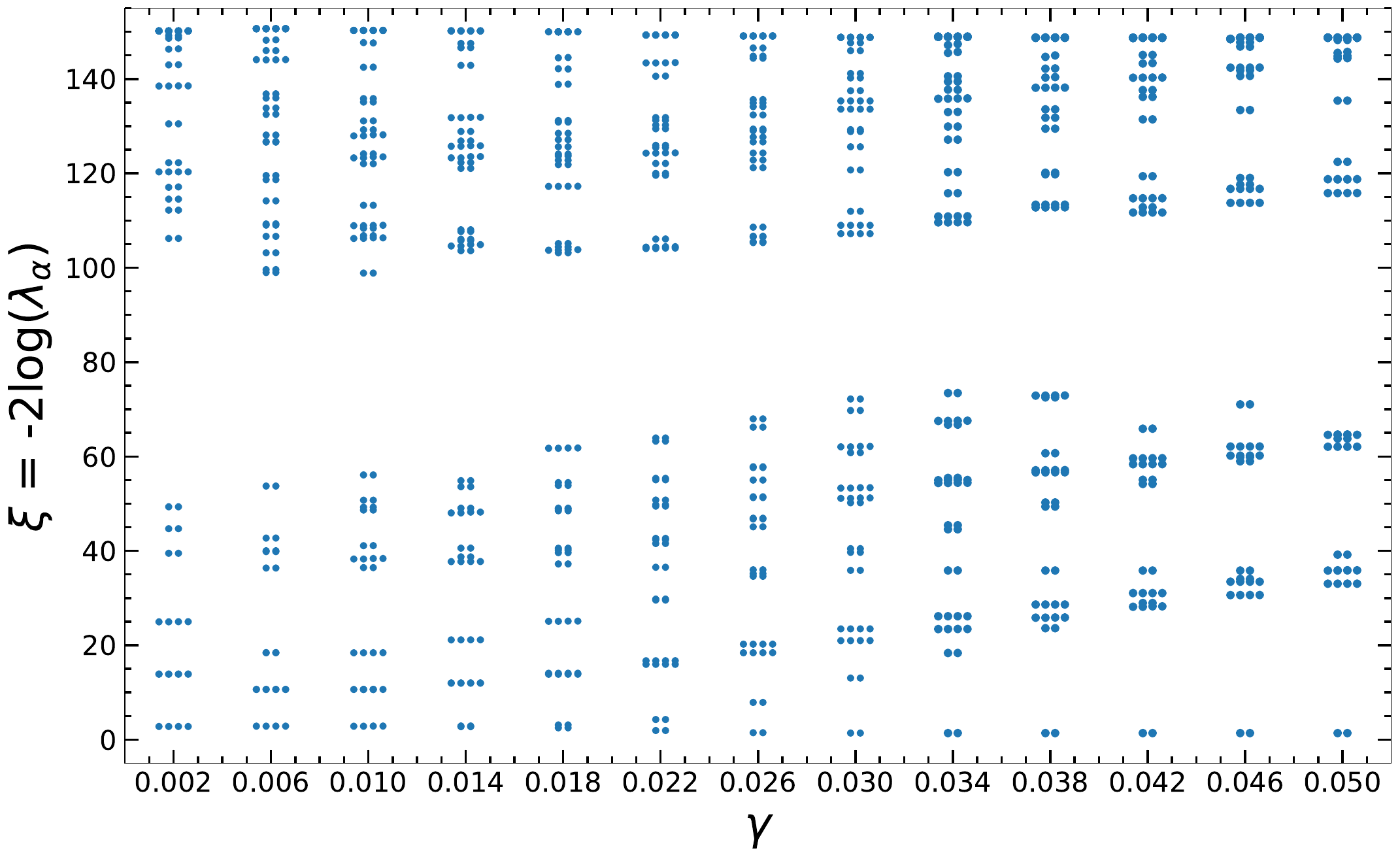}
    \caption{[Color online] Lowest part of the spectrum $\xi_\alpha$ for symmetric dissipation $L_\mu = S_z^2$, as a function of strength dissipation $\gamma$. The spectrum is organized in doublets, no matter the dissipation strength.}
    \label{figB2}
\end{figure}

In this work, we implement the technique proposed by Zwolak and Vidal \cite{Zwolak} based on MPOs, in the thermodynamic limit. The method allows us to simulate the time evolution driven by the master equation, and to evaluate the properties of the steady state. The algorithm consists of using, essentially, the TEBD algorithm for real-time simulation of Markovian dynamics, representing the mixed state $\rho$ as a pure state $\ket{\rho}_{\sharp}$ via Choi isomorphism $\rho = \sum_i p_i \ket{\Psi_i} \bra{\Psi_i} \rightarrow  \ket{\rho}_{\sharp} = \sum_i p_i \ket{\Psi_i} \otimes \ket{\Psi_i}$, see Fig.\ref{fig0}. We can then obtain the vectorized form of the Lindbladian super-operator as
\begin{eqnarray} \label{vLind}
    \mathcal{L}_\sharp &\equiv& - \left( H \otimes \mathbb{I}  - \mathbb{I} \otimes H^{T} \right)  \nonumber \\ &+& \sum_{\mu} \left( L_{\mu} \otimes L^{*}_{\mu} - \frac{1}{2} L^{\dag}_{\mu}L_{\mu} \otimes \mathbb{I} - \frac{1}{2} \mathbb{I} \otimes L^{*}_{\mu}L^{T}_{\mu} \right). 
\end{eqnarray}
The solution of the master equation can then be written as $\ket{\rho(T)}_\sharp = e^{T \mathcal{L}_\sharp} \ket{\rho(0)}$, and the steady state of the system is given by $\ket{\rho_s}_\sharp \equiv \lim_{T \rightarrow \infty} \ket{\rho(T)}_\sharp$. The density matrix $\ket{\rho_s}_\sharp$ is also the eigenvector of $\mathcal{L}_\sharp$ corresponding to zero eigenvalue, so that $\mathcal{L}_\sharp \ket{\rho_s}_\sharp = 0$. In our case, the Lindbladian super-operator can be written as a sum of local nearest-neighbor terms $\mathcal{L}[\rho] = \sum_i \mathcal{L}_{i, i+1}[\rho]$, so that we can implement the infinite-TEBD algorithm  \cite{OrusiTEBD, Vidal2} via a Trotter-Suzuki decomposition of the time evolution operator driven by the Lindbladian super-operator.

In the present study, {we consider the bond-dimension $\chi = 10$ in preparation of the initial pure state $\ket{\Psi}$. The MPO density matrix $\rho$ was computed with 4-rank $\Gamma$ tensors and dimensions $(\chi^2,d,\chi^2,d)$, with $d$ the local Hilbert space dimension. The vectorized MPO density matrix $\ket{\rho}_\sharp$  was written as a MPS with 3-rank $\Gamma$ tensors with dimensions $\left( \chi^2, d^2, \chi^2 \right)$. Also, we implemented a second-order Trotter-Suzuki decomposition of the time-evolution operator with time steps as small as $dt = 10^{-2}$ up to maximum time evolution $T = 300$. The above setup was sufficiently accurate, } allowing us to obtain the steady state in the thermodynamic limit. We used the quantity $\Delta \equiv$  $_\sharp$$\bra{\rho_s}  \mathcal{L}_\sharp \ket{\rho_s}_\sharp$ as convergence criterion which should tend to zero in time, meaning that we approach the steady state. We obtained values close enough to zero, with negligible imaginary part Im$(\Delta)$ and real part Re$(\Delta)$ smaller than $10^{-4}$. Furthermore, to assess the validity of our simulations we also computed the negative eigenvalues of the n-site reduced density matrix $\rho_n$, since the method itself does not guarantee a completely  positive density matrix. To be more precise, we computed the sum of the negative eigenvalues of $\rho_n$ and explicitly tested that this was always very close to zero.

\vspace{3pt}

{\it Results.-} We investigate the effects of dissipation on 1d SPT order by computing the Markovian dynamics of the AKLT model with Lindbladian jump operators $L_\mu$ that may respect or not the relevant symmetries protecting the SPT Haldane phase. {In the symmetric case, we consider a dissipation represented by local jump operators $L_i$ which satisfy the strong symmetry conditions discussed in Ref.\cite{oSPTP}. On the other hand, the non-symmetric case does not satisfy the symmetry condition}. We consider two cases of dissipation; first, we implement the non-symmetric dissipation $L_\mu = S_z$, which breaks time-reversal symmetry in the Liouvillian. Second, we consider the symmetric dissipation $L_\mu = S_z^2$, which preserves  time-reversal symmetry. In both cases, we compute the steady state of the system for different dissipation strengths $\gamma$, and evaluate the string order parameter $\mathcal{O}^{\alpha}_{S}$, the purity of the mixed state of $n$ consecutive sites $\Gamma_n \equiv {\rm tr}(\rho^2_n)$ ($< 1$ for mixed states), and the degeneracies present in the spectrum $\xi_\alpha$ of the Schmidt coefficients of steady state $\ket{\rho_s}_\sharp$.  

In the case of symmetry-breaking dissipation, $L_\mu = S_z$, we start from the AKLT state and compute the steady state under dissipation for $\gamma$ up to $0.05$ which, as we shall see, is already sufficiently large for our purposes. The string order parameter $\mathcal{O^{\alpha}_S}$ for $\alpha = x,y,z,z^2$ and purity $\Gamma_n$ are shown in Fig.\ref{figA1}. In the plots, we see that magnetic  and string order decays quickly with the dissipation strength, and becomes negligible for $\gamma > 0.02$, signaling that the SPT order initially present in the Haldane phase of the AKLT state does not remain under this type of dissipation. In fact, our calculations are also compatible with a pure steady state, since the purity goes also to one quickly. Additionally, the spectrum $\xi_\alpha$ is shown in Fig.\ref{figA2} for the different values of dissipation strength $\gamma$. As we can see in the figure, the spectrum does not show the typical degeneracy patterns of SPT order for any value of $\gamma$. All these observations show that SPT order is destroyed by this type of non-symmetric dissipation. This result is in agreement with previous investigations of the stability of symmetry-protected topological phases for non-interacting topological insulators coupled to a time-reversal breaking environment \cite{NatTR}.

Next, we consider the case of symmetric dissipation $L_\mu = S_z^2$. The results for string order parameters and purity are shown in Fig.\ref{figB1}. As we can see in the plot, string orders $\mathcal{O}^{x}_S$ and $\mathcal{O}^{y}_S$ drop equally to zero, whereas string orders $\mathcal{O}^{z}_S$ and $\mathcal{O}^{z^2}_S$ increase with the dissipation strength up to their maximum value $1$ for strong dissipation $\gamma > 0.02$. The non-zero value of these string order parameters signals the presence of non-trivial long-range behavior in the steady state, in agreement with recent discussions on the resilience of certain types of topological order for mixed states \cite{oSPTP}. {It's important to mention the value of $\mathcal{O}^{z}_S$ at the non-equilibrium steady state is different for the typical $\mathcal{O}^{\alpha}_S = \frac{4}{9}$ for pure AKLT state, which indicates that this type o symmetric dissipation $L_\mu = S_z^2$ leads to a different type of SPT order.} In addition, we can see that the purity flows towards $0.5$ for strong dissipation, meaning that we have mixed steady states. The overall picture is completed by assessing the spectrum $\xi_\alpha$, which we show in Fig.\ref{figB2}. Importantly, the spectrum, in this case, \emph{is always organized in terms of doublets, no matter the strength of the dissipation}, in analogy to the typical degeneracy of the entanglement spectrum of the Haldane SPT phase in the pure-sate case. In combination with the non-zero value of some string-order parameters, we take this as an indication that the symmetry-preserving dissipation indeed keeps some kind of long-range correlation order in the mixed steady state, being this the mixed-state analogous of SPT order in pure states. For all practical purposes, we can consider this as a realization of 1d dissipative SPT order, and it admits the same mathematical treatment as 1d SPT order for pure states in terms of irreducible representations of projective symmetry groups in the underlying tensor network.

\vspace{3pt}

{\it Conclusions and outlook.-} In this paper we have studied the effects of different types of dissipation on 1d SPT order. In particular, we have analyzed the steady states obtained by acting with time-reversal breaking and non-breaking dissipation on the Haldane phase of the 1d AKLT model. We have seen that non-symmetric dissipation breaks all SPT signatures already for very small values of the dissipation strength, {resulting in reduced density matrices representing pure local states. It would be interesting to check if this effect remains for more general multi-site Lindblad operators.} Symmetric dissipation induces a long-range correlation order in the mixed steady state, that is analogous to the 1d SPT order in pure states. This has been assessed by the non-zero values of some string order parameters, as well as by the degeneracy of the spectrum of Schmidt coefficients in the MPO representation of the mixes state, akin to the degeneracy of the entanglement spectrum in 1d SPT phases. Our results point towards a broad concept of 1d dissipative SPT order, which can have the same mathematical treatment as 1d SPT order for pure states, in terms of irreducible representations of projective symmetry groups. This opens the possibility to investigate new dissipative phases of quantum matter, where non-local patterns of long-range order are at play, and which can be naturally described and characterized using the tensor network language, probably even beyond 1d systems. Diving into such an analysis is a topic for future works. 

\bigskip

{\bf Acknowledgements:} This work was supported by DIPC (Donostia International Physics Center), CNPq(Conselho Nacional de Desenvolvimento Científico e Tecnológico), and CAPES (Coordenação de Aperfeiçoamento de Pessoal de Nivel Superior). L. M. V. acknowledges CAPES and DIPC for all support in the joint visiting researcher program. R. O. acknowledges DIPC, Ikerbasque, Basque Government, and Diputaci\'on de Gipuzkoa. 

\medskip

\bibliographystyle{ieeetr}
\typeout{}
\bibliography{library}

\end{document}